\documentclass[traditabstract]{aa} 
\usepackage{graphicx}
\usepackage{txfonts}
\usepackage{natbib}

\begin{document}
   \title{Cold gas in the inner regions of intermediate redshift clusters \thanks{Based on observations carried out with the IRAM Plateau de Bure Interferometer. IRAM is supported by INSU/CNRS (France), MPG (Germany) and IGN (Spain).}
}

   \author{P. Jablonka
          \inst{1,2}
          \and
          F. Combes\inst{3}
          \and
	  K. Rines \inst{4}
          \and
	  R. Finn \inst{5}
          \and
          T. Welch \inst{4}          
          }

   \institute{Laboratoire d'Astrophysique, Ecole Polytechnique F\'ed\'erale de Lausanne (EPFL), 1290, Sauverny, Switzerland 
	\and 
	GEPI, Observatoire de Paris, CNRS UMR 8111, Universit\'e Paris Diderot, 92125, Meudon Cedex, France 
              \email{pascale.jablonka@epfl.ch}
         \and
             Observatoire de Paris, LERMA \& CNRS: UMR8112, 61 Av. de l'Observatoire, 75014, Paris
	\and
	Department of Physics and Astronomy, Western Washington University, Bellingham, WA 98225, USA    
         \and 
	Department of Physics, Siena College, 515 Loudon Road, Loudonville, NY 12211, USA
             }

   \date{Received ; accepted}

 \abstract{
Determining the nature and modes of star formation at galactic scales requires an
understanding of the relationship between the gas content of a galaxy and its star
formation rate. Remarkable progress has been made in understanding the conversion
mechanisms in field galaxies, but the cold and dense gas fueling the star formation
has been much less investigated in galaxies inside clusters. We present the first CO
observations of luminous infrared galaxies (LIRGs) inside the virial radii of two
intermediate redshift clusters, CL1416+4446 (z=0.397) and CL0926+1242 (z=0.489). We
detect three galaxies at high significance (5 to 10 $\sigma$), and provide robust estimates
of their CO luminosities, L$^{\prime}$$_{\mathrm{CO}}$.  In order to put our results into a general context,
we revisit the relation between cold and hot gas and stellar mass in nearby field and
cluster galaxies. We find evidence that at fixed L$_{\mathrm{IR}}$ (or fixed stellar mass), the
frequency of high L$^{\prime}$$_{\mathrm{CO}}$ galaxies is lower in clusters than in the field, suggesting
environmental depletion of the reservoir of cold gas. The level of star formation
activity in a galaxy is primarily linked to the amount of cold gas, rather than to
the galaxy mass or the lookback time. In clusters, just as in the field, the
conversion between gas and stars seems universal. The relation between L$_{\mathrm{IR}}$ and L$^{\prime}$$_{\mathrm{CO}}$
for distant cluster galaxies extends the relation of nearby galaxies to higher IR
luminosities. Nevertheless, the intermediate redshift galaxies fall well within the
dispersion of the trend defined by local systems. Considering that L$^{\prime}$$_{\mathrm{CO}}$ is generally
derived from the CO(1-0) line and sensitive to the vast majority of the molecular gas
in the cold interstellar medium of galaxies, but less to the part which will actually
be used to form stars, we suggest that molecular gas can be stripped before the star
formation rate is affected. Combining the sample of Geach et al. (2009, 2011) and
ours, we find evidence for a decrease in CO towards the cluster centers. This is the
first hint of an environmental impact on cold gas at intermediate redshift.
}

  \maketitle

%

\section{Introduction}

It has long been known from optical observations that galaxy properties
change with time, and that they also change systematically with the galaxy environment.
Observed trends include morphological transformations
\citep[e.g.,][]{dressler1980, desai2007} and signatures of star formation
such as broadband colors \citep[e.g.,][]{butcheroemler1984}.   Dusty star forming galaxies, best
investigated in the infrared, contribute significantly more to the integrated
cluster star formation rate (SFR) than do the optically selected blue galaxies.  For
example, the SFRs derived from [OII] are 10 to 100 times lower than the rates
estimated from IR luminosities \citep[e.g.,][]{duc2002, geach2009a}.  Because
clusters hardly contain any ULIRGs, galaxy evolution must be traced with LIRGs
\citep[Log(LIR/L$_{\odot}$) $>$ 11;][]{finn2010, wardlow2010}. From {\it Spitzer}
MIPS 24 $\mu$m data, \citet{saintonge2008} and \citet{finn2010} could establish
that the fraction of LIRGs declines exponentially with time both in the field
and in clusters, however the LIRG fraction in clusters lies significantly below the
fraction in the field at all epochs. \citet{haines2009} found that intermediate redshift
LIRGs are preferentially
located at large cluster-centric radii; they interpret their observations as a
combination of both the global decline in star formation in the universe since
z$\sim$1 and enhanced star formation in the infall regions of clusters.

The relationship between gas content and star formation rate is necessary to
understand the nature and modes of star formation at
the galactic scale.  Remarkable progress has been made
toward understanding the conversion mechanisms in field galaxies
\citep{wong2002,bigiel2008, daddi2010, genzel2010, tacconi2010,combes2011}. 

Conversely, the cold and dense gas fueling the star formation has hardly been
investigated in clusters, despite their exceptional opportunities for cosmological
studies. The Coma supercluster and Virgo Cluster are the only sites of investigation
\citep[e.g.,][]{kenney1989, Rengarajan1992, boselli1997, casoli1998,
  lavezzi1998,scott2013}.  For the largest sample of more distant cluster
galaxies, \citet{geach2009, geach2011} measured CO in the outskirts (1.5$-$3
$\times$ R$_{200}$ where $R_{200}$ encloses an average density of 200 times
the critical density and roughly corresponds to the virial radius) of
CL0024+16 ($\sigma$=911km/s, z=0.395).  Notably, no CO
measurements existed {\it inside} the virial radius of distant
clusters before now.   Observations of galaxies covering a wide radial range
can help distinguish  the various physical processes expected to play a role,
e.g. ram-pressure stripping (interaction with the intergalactic medium),
strangulation (removal of any envelope of gas), tidal interactions, and mergers,
because these processes peak in effectiveness at different clustercentric radii \citep{delucia2010,
  moran2007}.  How, when, and where they affect the evolution of galaxies has
still to be explored.

This paper reports on the first CO observations of galaxies located within
R$_{200}$ of two intermediate redshift clusters. Section \ref{local} starts by
synthesizing the comparison between local field and cluster galaxies. Section
\ref{intermediate} presents our observations obtained with the IRAM PdBI, while
Sect. \ref{results} discusses our results. We summarize our work in
Sect. \ref{conclusion}.


\section{Where do we stand ? The case of the local samples}
\label{local}

Historically, the question of the effect of the environment onto star forming
galaxies has been framed in terms of a possible correlation between an observed
depletion in HI and the corresponding lower CO content in cluster galaxies.
\citet{haynes1984} showed a strong correlation between the HI mass of spiral galaxies
and their optical size, independent of their exact Hubble type. This correlation is
interpreted as providing the normal neutral gas content of a galaxy in the absence of
environmental effects. Any difference between the above fit and the observed mass of
HI in a galaxy is called deficiency (although note that the dispersion around the
mean trend is not negligible).

In the challenging route to understanding whether molecular gas can be stripped in
a similar way HI is, there are con- and pro-, all based on studies in the local
Universe. The persistence of publications on this topic suggests that one has not yet
convincingly solved and/or quantified the question either way.

\citet{kenney1989} and later \citet{Rengarajan1992} analyzed the molecular and
atomic gas properties of galaxies in Virgo and found evidence for stripping in CO,
although at lower rate than in HI (2 versus 10 in deficiency). The Coma
supercluster has given rise to a long series of studies, several of which report
no difference in CO content between field and supercluster galaxies nor any
variation linked to HI \citep{casoli1991, casoli1996, casoli1998,boselli1997}.
However, recent evidence for ram-pressure stripping of CO was found in Virgo
  and in Abell 1367, one cluster of the Coma supercluster \citep{vollmer2008,
    scott2013}, although the frequency of this phenomenon seems low
  \citep{vollmer2012}.  \citet{lavezzi1998} investigated 11 nearby clusters and
found no variation of the ratio M(H$_2$)/M(HI) with clustercentric distance,
whereas the L$_{\mathrm{FIR}}$/ M(H$_2$) decreases towards the cluster center,
suggesting higher star formation efficiency at the cluster periphery.  However
their distances were not scaled to the cluster structural parameters (e.g., the
virial radius) that could play a significant role given the large dispersion in
cluster properties (e.g., velocity dispersions ranging from $\sim$1200km/s to
$\sim$300km/s). It is difficult to reconcile the above discrepant conclusions.
Possible explanations include small sample sizes, different levels of star
formation activity, differing statistical methods, and questionable normalization
of the gas masses (e.g., D$_{25}$ works well for HI but is less secure for H$_2$).

In the following, we attempt to determine environmental effects by
focusing on the link between the intensity of star
formation (as traced by the galaxy total far infrared luminosity, L$_{\mathrm{IR}}$)
and the galaxy reservoir of molecular gas (as indicated by the luminosity in CO,
L$^{\prime}${$_{\mathrm{CO}}$). 

We compiled datasets from the literature with IR and CO observations.  We restricted
our analysis to rather massive galaxies (L$_{\mathrm{IR}} \ge 10^9$ L$_{\odot}$ ) in
which the metallicity is sufficiently high to ensure that CO reliably traces
H$_2$. Our field sample comprises the FCRAO Extragalactic CO Survey,
log(L$_{\mathrm{IR}}$/L$_{\odot}$) $\in$[8.9,11.6], z$_{\mathrm{median}}$=0.0046
\citep{young1995}, the Galex Arecibo SDSS Survey (GASS),
log(L$_{\mathrm{IR}}$/L$_{\odot}$) $\in$[9.7,11.2], z$_{\mathrm{median}}$=0.03,
\citep{saintonge2011}, the Nobeyama CO Atlas of Nearby Spiral Galaxies,
log(L$_{\mathrm{IR}}$/L$_{\odot}$) $\in$[9.2,11.2], z$_{\mathrm{median}}$=0.002,
\citep{kuno2007}, the BIMA survey, log(L$_{\mathrm{IR}}$/L$_{\odot}$) $\in$[8., 11.2],
z$_{\mathrm{median}}$=0.002, \citep{helfer2003}, and the observations of
\citet{gao2004}, log(L$_{\mathrm{IR}}$/L$_{\odot}$) $\in$[9.9,12.5],
z$_{\mathrm{median}}$=0.007, and \citet{garciaburillo2012},
log(L$_{\mathrm{IR}}$/L$_{\odot}$) $\in$[11.,11.9], z$_{\mathrm{median}}$=0.015,
whereas for our cluster galaxies, we gathered the galaxy samples of
\citet{kenney1989} in Virgo, log(L$_{\mathrm{IR}}$/L$_{\odot}$) $\in$[9.,10.3],
\citet{lavezzi1999}, log(L$_{\mathrm{IR}}$/L$_{\odot}$) $\in$[10.,11.3]}, and
\citet{casoli1991}, log(L$_{\mathrm{IR}}$/L$_{\odot}$) $\in$[9.9,11.1],in Coma, the
collection of clusters and groups of the Pisce-Perseus and the Coma/A1367 super
clusters from \citet{lavezzi1998, boselli1997, scott2013},
log(L$_{\mathrm{IR}}$/L$_{\odot}$) $\in$[9.,11.]. From the GASS sample, we
distinguished galaxies in groups (defined as $>$ 10 members at the same spectroscopic
redshift) from galaxies in isolation using the work of \citet{tempel2012}. Similarly
the Nobeyama survey encompasses galaxies located in Virgo, and the BIMA survey have
some in the Ursa Major cluster, which were added to the cluster sample, and finally
the isolated galaxies of \citet{boselli1997} were attached to our field sample.  In
total, we gathered 275 field and 170 cluster galaxies for which we could collect both
reliable far infrared fluxes and CO luminosities.

For the sake of homogeneity, we revised the distances of all galaxies: we use the
redshift-independent indicators provided in NED for the most nearby galaxies and
we calculate the luminosity distance from the redshifts of the remaining galaxies
(assuming $\Omega_M$=0.3, $\Omega_\Lambda$= 0.7, h =0.70). For all samples except GASS, the far
infrared fluxes have been derived from the {\it IRAS} bands as follows:

F$_{\mathrm{IR}}$=1.8$\times$10$^{14}$(13.48f$_{12}$ + 5.16f$_{25}$ + 2.58f$_{60}$
+ f$_{100}$) [W.m$^{-2}$], \citep{sanders1996}.  L$^{\prime}${$_{\mathrm{CO}}$ is
  defined as in \citep{SolomonVandenBout}.

For the GASS sample, we restrict our analysis to the 164 galaxies with measured CO fluxes
(discarding those with only upper limits). 
Out of this sample, 26 galaxies have measured {\it IRAS} IR fluxes, while 102 galaxies
have {\it WISE} 22$\mu$m
observations.  We derive the total IR luminosity of the WISE-detected galaxies
using their mid-IR fluxes, as measured in apertures corresponding to their 2MASS
major axes. \citet{jarret2013} showed that fluxes measured in
the {\it WISE} 22$\mu$m and in the {\it Spitzer} 24$\mu$m bands are equivalent. Therefore, we
used the relation of \citet{rieke2009} between L(24) and L(TIR). 
We note that for all galaxies with both {\it WISE } and {\it  IRAS} IR luminosities (both
the GASS sample and the other studies described above), there is a one to
one correspondence (with small dispersion) between the IR luminosity estimates.

When necessary (e.g., when the antenna beam sizes were smaller than the galaxies and
no mapping is available), the CO fluxes were aperture-corrected following
\cite{young1995}, i.e., assuming an exponential spatial distribution of CO, with a
scalelength of a tenth of D$_{25}$, the galaxy diameter.  The mean of the
  corrective factors is 1.17, with a median at 1.08 and a maximum at 1.8.

We intentionally report our results in terms of luminosity rather than $H_2$ masses
and star formation rates, in order to free ourselves from the uncertainties in
conversion factors.

\begin{figure} 
\resizebox{\hsize}{!}{\includegraphics[angle=0]{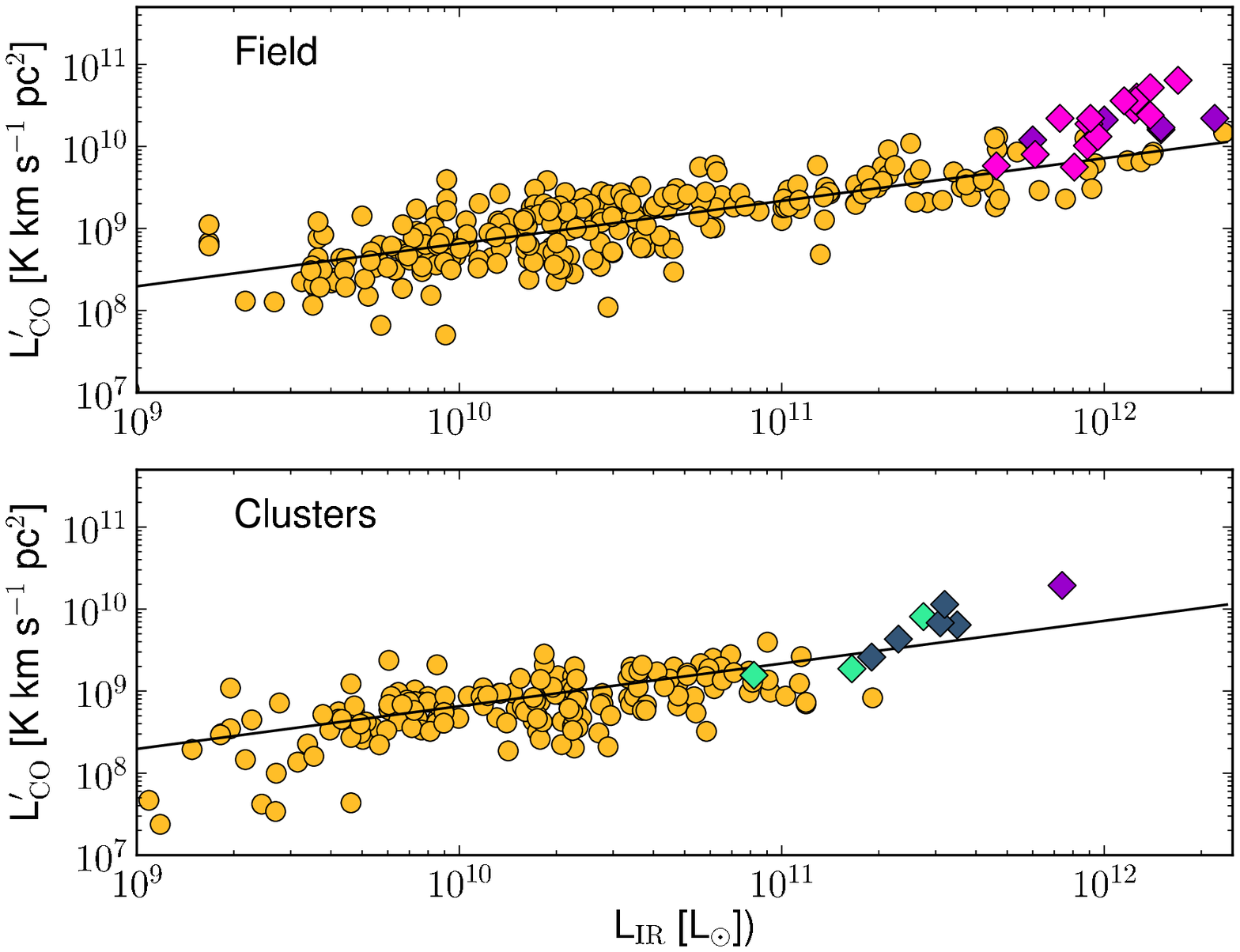}}
\caption{The relation between L$_{\mathrm{IR}}$ and L$^{\prime}$$_{\mathrm{CO}}$ for
  field (upper panel) and cluster (lower panel) galaxies. The orange circles show the
  sample of nearby galaxies defined in Section \ref{local}, green rhombuses show
  intermediate redshift cluster galaxies, from this work (light) and
  \citet{geach2009, geach2011} (dark). The purple circles show z$>1$ galaxies in
  clusters \citep{wagg2012} and in the field \citep{daddi2010}. Pink identifies the 
  \citet{tacconi2010}'s sample of normal galaxies  at  z$\sim$1.4 and z$\sim$2.
  In both panels, the  line shows the fit of the L$^{\prime}$$_{\mathrm{CO}}$- L$_{\mathrm{IR}}$ relation
  for the field galaxies over the full IR luminosity range, L$_{\odot}$. log$_{10}$(L$^{\prime}$$_{\mathrm{CO}}$) =
     0.52$\times$log$_{10}$(L$_{\mathrm{IR}}$) + 3.61 .}
\label{LIR-LCO}
\end{figure}

 Before describing our new observations of galaxies in moderate-redshift clusters, we
 first compare field and cluster galaxies at low redshift.  Figure \ref{LIR-LCO}
 displays the relation between L$_{\mathrm{IR}}$, and L$^{\prime}${$_{\mathrm{CO}}$
   for our compilation of field and cluster galaxies. We fit the relation between
   L$_{\mathrm{IR}}$ and L$^{\prime}${$_{\mathrm{CO}}$} over the full range
   10$^9$-10$^{12}$: L$_{\odot}$. log$_{10}$(L$^{\prime}$$_{\mathrm{CO}}$) =
     0.52$\times$log$_{10}$(L$_{\mathrm{IR}}$) + 3.61, very close to the relation of
   \citep{SolomonVandenBout}.

  At fixed L$_{\mathrm{IR}}$, the cluster galaxies are
   down-shifted compared to the system in the field, suggesting that for a given star
   formation rate, the reservoir of cold gas is smaller in cluster galaxies.

\begin{figure}
\resizebox{\hsize}{!}{\includegraphics[angle=0]{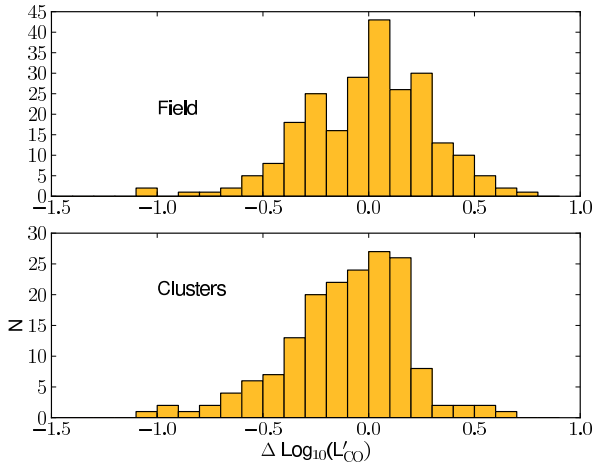}}
\caption{The residuals (in log scale) of the difference between the observed CO luminosity and the
fit of the relation between L$_{\mathrm{IR}}$ and L$^{\prime}$$_{\mathrm{CO}}$ for field galaxies and L$_{\mathrm{IR}}$
between 10$^9$ and 10$^{12}$ L$_{\odot}$. We use only local galaxies for the comparison.}
\label{histo}
\end{figure}

Figure \ref{histo} shows the residuals of the
observed CO luminosities from the best-fit
L$_{\mathrm{IR}}$-L$^{\prime}${$_{\mathrm{CO}}$ relation of field galaxies
(restricted to the 237 galaxies in with infrared luminosities between 10$^9$ and
  10$^{12}$ L$_{\odot}$). Figure
  \ref{histo}  illustrates the difference between the cluster and field samples: while
  field galaxies can have CO luminosities as small as cluster galaxies, the latter
  are essentially never as luminous in CO as field galaxies with similar infrared
  luminosity. A Kolmogorov-Smirnov test rejects the null hypothesis (samples are
  drawn from the same distribution) with a probability (p-value) of much less
  than 1\% that the null hypothesis is valid.

We estimate the stellar masses (M$_{\star}$) of 231 field and 137 cluster nearby
galaxies with available SDSS colors and/or 2MASS J, H and K-bands data using the code
Fitting and Assessment of Synthetic Templates \citep[FAST,][]{kriek2009}. We assumed
a Chabrier IMF, a solar metallicity, an exponentially decaying star formation rate
and the \citet{bruzual2003} synthesis models.

Fig. \ref{LCOMstar} shows the relation between stellar mass and CO luminosity.  The
fit of the relation between L$^{\prime}$$_{\mathrm{CO}}$ and M$_{\star}$ for the
nearby field galaxies is shown in both panels. It is performed on galaxies with
L$_{\mathrm{IR}}$$<$ 2$\times$10$^{11}$L$_{\odot}$ to allow comparison between the
clusters and the field.  log$_{10}$(L$^{\prime}$$_{\mathrm{CO}}$) =
  0.37$\times$log$_{10}$(M$_{\star}$) + 5.04.

There is more scatter in the relation between mass and CO luminosity  than
  between L$_{\mathrm{IR}}$ and L$^{\prime}$$_{\mathrm{CO}}$. 
We divide galaxies in three L$_{\mathrm{IR}}$ ranges. The first threshold,
2$\times$10$^{10}$L$_{\odot}$, closely corresponds to the median L$_{\mathrm{IR}}$ of
the nearby field galaxies, while the second threshold, 5$\times$10$^{11}$L$_{\odot}$,
correspond to the lowest L$_{\mathrm{IR}}$ of the distant field galaxies.  In an
independent way, Fig. \ref{LCOMstar} confirms that CO luminous galaxies are more
common in the field than in clusters.
 Only 7\% of the cluster galaxies versus 20\% of the field galaxies are
found 0.3dex above the fit line. Moreover, as indicated by the clear segregation of
galaxies as a function of L$_{\mathrm{IR}}$, one sees that the level of star formation
activity is primarily linked to the amount of cold gas, rather than to the galaxy
mass or redshift. Genzel et al. (2010) reached similar conclusions when they
discussed a universal gas/star formation relation.

\begin{figure} 
\resizebox{\hsize}{!}{\includegraphics[angle=0]{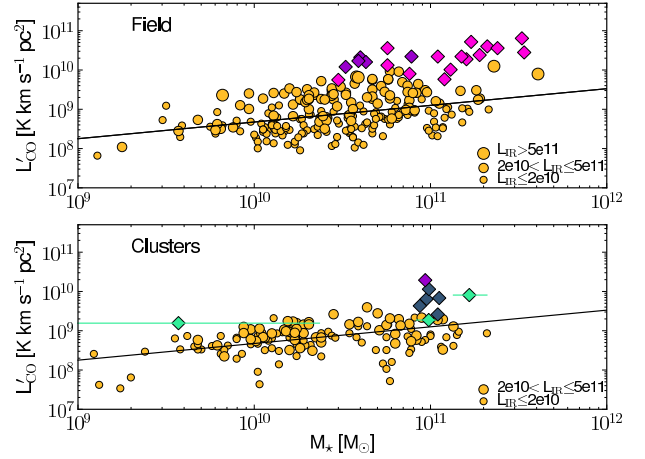}}
\caption{The relation between the galaxy stellar mass and its CO luminosity. The
  color code is the same as in Figure \ref{LIR-LCO}.  In both panels the straight line
corresponds to the fit of the relation for the field galaxies  with L$_{\mathrm{IR}}$$<$ 2$\times$10$^{11}$ L$_{\odot}$
to allow comparison with the clusters. We show the uncertainties of the mass estimates for our new sample
of cluster galaxies.}
\label{LCOMstar}
\end{figure}

There are a few caveats to this result. First, although we attempted to homogenize
all quantities (distances, aperture correction, luminosities, masses), our
compilation is drawn from datasets acquired on from different telescopes and
techniques across several years. Only two studies \citep{young1995,gao2004} span
nearly the full range of infrared luminosities, and only the study of
\citet{kuno2007} contains enough galaxies in clusters and in the field allowing a
direct comparison from identical observational conditions.  Second, some galaxies do
not have complete spatial CO coverage. Although in principle the aperture corrections
are small, the exact form and the magnitude of this correction can significantly vary
from one galaxy to another \citep{saintonge2011}.  Finally, optical parameters such
as diameters may depend on what wavelength they are measured at and the details of
the observations (e.g., CCD photometry versus photographic plates).  Nevertheless,
while systematics could affect the comparison between the samples, there is no
evidence that these would affect the clusters more than the field.

Unfortunately, most of the nearby galaxies are at the latest stage of their evolution
and the lowest point of their star forming activity: \citet{finn2010} and
\citet{saintonge2008} have shown that the fraction of LIRGs declines exponentially
with time both in the field and in clusters, hence hindering our ability to identify
where and when the environment affects the evolution of the galaxies.  This motivated
our observing programme of intermediate redshift LIRGs.

\section{Intermediate redshift clusters}
\label{intermediate}
\subsection{The sample}

One obstacle to determining the impact of environment on star formation
at moderate redshift is that few cluster galaxies are classified as LIRGs or
ULIRGs \citep[e.g.,][]{finn2010}.  We identified new cluster LIRGs among
{\em Spitzer} and MMT/Hectospec observations of X-ray-selected clusters
from the 400 square degree serendipitous survey \citep{burenin2007} that
are included in the
{\em Chandra} Cluster Cosmology Project sample \citep{vikhlinin2009}.
We analyzed {\it Spitzer} IRAC and MIPS observations of several clusters
at $z\geq0.35$ to identify candidate LIRGs.  We then use MMT/Hectospec
optical spectroscopy (Rines et al.~in preparation) to classify the LIRGs
as cluster members or background galaxies. 

For the pilot program described here, we selected two LIRGs LIRGs that lie inside the
radius R$_{200}$ of their clusters.  

One target, GAL1416+4446, lies in CL1416+4446 ($z$=0.397, projected velocity
dispersion $\sigma_p \approx$750 km/s, M$_{500}$=$(2.52\pm0.24)\times 10^{14}$
M$_{\odot}$), at 0.82R$_{200}$, with z$_{spec}=0.3964 \pm 0.00008$.  The second
target, GAL0926+1242, was found in CL0926+1242 ($z$=0.489, $\sigma_p \approx$810
km/s, M$_{500}$=$(3.00\pm0.30)\times 10^{14}$M$_{\odot}$), at 0.38R$_{200}$, with
z$_{spec}=0.4886 \pm 0.0001$. We estimate R$_{200}$ by scaling R$_{500}$ from the
{em Chandra} mass estimate assuming an NFW profile \citep{navarro1997} with $c$=5.

The projected radii of the galaxies may not fully reflect the genuine 3D
  positions of the galaxies. Though, GAL1416+4446 and GAL0926+1242-A/B have a fair to
  very high probability to be indeed located inside the virial radius of their parent
  clusters. \citet{mahajan2011} break down the 3D locations of
  galaxies in and around a simulated cluster by both projected radius and projected
  velocity offset (their Table 2). GAL1416+4446 (0.5R$_{200}$ $\le$ r $\le$
  R$_{200}$, $\delta$v$_{\mathrm{LOS}}$/$\sigma_{\mathrm{cluster}}$ $\sim 0.17$) fall
  in the bins for which 49\% of galaxies have 3D radii within the cluster virial
  radius, while an additional 18\% are part of the so-called ``backsplash" population, i.e.,
  galaxies that have passed inside R$_{200}$ but are now further away.
  GAL0926+1242-A and -B (0$\le$ r $\le$0.5R$_{200}$,
  $\delta$v$_{\mathrm{LOS}}$/$\sigma_{\mathrm{cluster}}$ $\sim 0.1$) fall in the bins
  for which 89\% of galaxies have 3D radii within the cluster virial radius.  Thus,
  it is most likely that GAL0926+1242-A and -B and GAL1416+4446 have been subject
  to environmental effects.

Both CL1416+4446 and CL0926+1242 were observed with {\it Spitzer}/IRAC and MIPS at
3.6, 4.5, 5.8, 8 and 24 $\mu$m. We used the 24 $\mu$m fluxes of our two targets
to estimate total infrared luminosities following the method of \citet{finn2010}.
We find  $L_{IR}=2.6\times 10^{11}$ L$_{\odot}$ (GAL1416+4446) and
$2.2\times 10^{11}$ L$_{\odot}$ (GAL0926+1242),
equivalent to star formation rates of 45 and 38 M$_{\odot}$/yr, respectively.
These galaxies are the brightest confirmed cluster members in the far infrared
among twelve CCCP clusters with $z$=0.35-0.50. These galaxies were the brightest
cluster members in the far infrared and the only ones at these flux levels.

We use two methods to exclude AGN activity as the source of the infrared emission in
these galaxies.  First, we determined that the IRAC colors of the galaxies are
consistent with dust heated by star formation rather than AGN activity
\citep{stern2005,sajina2005,pope2008}. Second, none of our sources are detected by
      {\em Chandra} (luminosity limits, L$_X$$>10^{42}$ erg/s; Alexey Vikhlinin,
      private communication).  Thus, the infrared luminosities of these
      spectroscopically confirmed members are dominated by star formation; they are
      neither X-ray sources nor mid-infrared AGN.

A close inspection of the CO map around GAL0926+1242 later revealed two actual
emission peaks (Fig. \ref{detection0926+1242}).  The Spitzer
Source List includes a companion to our primary galaxy target, which
was detected at shorter infrared wavelengths with IRAC. In the following text and
in Tables \ref{table:positions}, \ref{table:photometry}, and \ref{table:results},
we shall refer to GAL0926+1242-A as our primary target and to GAL0926+1242-B as
the secondary serendipitous CO detection. Table \ref{table:positions}
provides the coordinates, redshifts and clustercentric distances of the three
galaxies. Table \ref{table:photometry} summarizes the far infrared properties of
our sample.  In order to derive the total infrared luminosities of the two systems
in CL 0926+124, we assumed the same flux ratios at 24 $\mu$m as seen at 8
$\mu$m (the two sources are blended at 24 $\mu$m).

\begin{table*}[t!]
\caption{Positions of our intermediate redshift targets. We provide  the coordinates as in the HST and {\it Spitzer} catalogs.
 d$_{\mathrm{center}}$ is the distance to the cluster center in unit of R$_{200}$.  }             
\label{table:positions}      
\begin{tabular}{l c c c | c c | c c} 
\hline
	         & & \multicolumn{2}{c}{Spitzer}  & \multicolumn{2}{c}{HST}     &      \\
ID & Spitzer ID          & RA        & DEC        & RA      & DEC       & z  & d$_{\mathrm{center}}$  \\     
                 &  \multicolumn{2}{c}{J(2000)} & \multicolumn{2}{c}{J(2000)}   & & (R$_{200})$\\
\hline\hline         	      
GAL1416+4446 & SSTSLP J141619.52+444357.1  & 214.081350  &   44.732533 &    &  & 0.3964  & 0.82\\
GAL0926+1242-A & SSTSLP J092632.23+124213.0  & 141.634318 & 12.703622 & 09:26:32.26 & 12:42:11.72 & 0.4886  & 0.38\\
GAL0926+1242-B& SSTSLP J092631.98+124212.5   & 141.633271 & 12.703498& 09:26:32.05  & 12:42:12.36 &  0.4886   &  0.39\\
\hline
\end{tabular}
\end{table*}

\begin{table*}[t!]
\caption{Spitzer Photometry for our intermediate redshift targets.}             
\label{table:photometry}      
\begin{tabular}{l c c c c c c  } 
\hline
    ID                     & L$_{\mathrm{IR}}$ &    F$_{24}$ & F$_{3.6}$ & F$_{4.5}$ & F$_{5.8}$ & F$_{8.0}$  \\
          		    & (10$^{11}$L$_{\odot}$)      &       \multicolumn{5}{c}{($\mu$Jy)} \\
\hline\hline 
GAL1416+4446  & 2.75 & $1537.48 \pm 26.79$ & $207.75 \pm 0.45$ & $232.72	\pm 0.66 $ & $202.21 \pm 1.72$ & $460.49 \pm 3.19 $\\
\vspace{0.4mm}
GAL0926+1242-A  & 1.65    & $747.35 \pm 18.18$ &   $70.92 \pm 0.27$ & $62.96 \pm 0.36$ & $60.02 \pm 1.35$ &  $132.53 \pm 2.42$ \\
GAL0926+1242-B  & 0.82  &                      &   $31.69 \pm 0.21$ & $30.33 \pm 0.31$ & $33.12 \pm 1.28$ &  $70.65 \pm 2.36$ \\
\hline
\end{tabular}
\end{table*}

\subsection{CO observations}

We observed the cluster LIRGs with the IRAM PdBI on 2011 July (18, 23, 24) (CL
0926+1242) and 2011 November 8-9 (CL 1416+4446). The exposure times were 8h on
both sources, using five antennae. The observing conditions were excellent in
terms of atmospheric phase stability; however, any anomalous and high phase-noise
visibilities were flagged.  We benefited from 6 additional hours (March 28 \& 29,
April 6 \& 9) in the form of director discretionary time for CL 0926+1242.

We calibrated, mapped and analysed the data using the IRAM GILDAS software
\citep{guilloteau2000}. For CL 1416+4446, we targeted the CO(1-0) 115.27GHz
rotational transition, which at z = 0.396 is redshifted into the 3-mm band, with
$\nu_{obs}$ = 82.546 GHz.  The half power primary beam (field of view) is 61.4
arcsec.  The observed configuration D provided a 7.5 x 6.5 arcsec$^2$ beam with PA=
175 deg.  The GILDAS reduction was made into a cube of 128 $\times$ 128 $\times$ 180
pixels, of 1.5 $\times$ 1.5 arcsec $\times$ 20 MHz, or 8 $\times$ 8kpc $\times$ 72.6
km/s.  The rms noise level was 0.7 mJy /beam.  The spatial resolution does not allow
to resolve the galaxy CO emission.  No continuum was detected, with an rms noise
level of 0.05 mJy/beam in a 3.6 GHz bandwidth.

For CL 0926+1242, we targeted the CO(2-1) line at 230.538001 GHz, redshifted at
$\nu_{obs}$ = 154.866GHz.  The half power primary beam (field of view) is 32.5
arcsec.  The observed configuration C provided a beam of 2.0 $\times$ 1.7 arcsec$^2$
with PA= 21 deg.  The GILDAS reduction was made into a cube of 256 $\times$ 256
$\times$ 90 pixels, of 0.45 $\times$ 0.45 arcsec $\times$ 40 MHz, or 2.7 $\times$
2.7kpc $\times$ 77.5 km/s.  The rms noise level was 0.7 mJy /beam.  The spatial
resolution allows to separate two interacting galaxies, but not to resolve the galaxy
CO emission. No continuum was detected, with an rms noise level of 0.07 mJy/beam in
3.6 GHz bandwidth.

Our resolution in frequency was smoothed to 20MHz for CO(1-0)
in  CL 1416+4446 (72.615 km/s per channel) and 40MHz for CO(2-1)
in CL 0926+1242 (77.432 km/s per channel).

\section{Results}
\label{results}

Figure \ref{detection0926+1242} shows the CO detection map and spectra of
the two sources (designated A and B) in the field of CL0926+1242.
Fig. \ref{hst-co-cl0926+1242} shows the HST ACS/WFC F814W image of the targeted field
\citep{hoekstra2011}, marking the positions of what turns out to be two
interacting spiral galaxies, on which we superimposed the contours of their CO
fluxes. Figure \ref{detection1416+4446} shows the detection map and
spectrum of GAL1416+4446. GAL0926+1242-A and B are
detected at $\sim$5$\sigma$, while GAL 1416+4446 is detected at $\sim$ 10$\sigma$.

\begin{figure}[t!]
\centerline{
\includegraphics[angle=-90, scale=0.4]{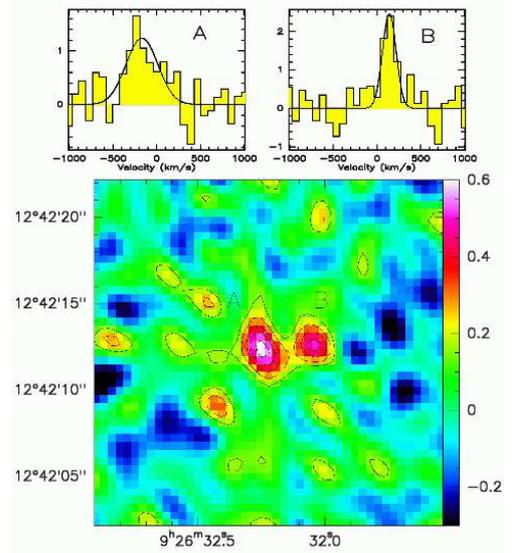}}
\caption{Intensity map and spectra of GAL0926+1242 A and B. The color wedge of the intensity map is in Jy km/s, while
the spectra display S$_{\mathrm{CO}}$  in mJy}
\label{detection0926+1242}
\end{figure}

\begin{figure}
\resizebox{\hsize}{!}{\includegraphics[angle=-90]{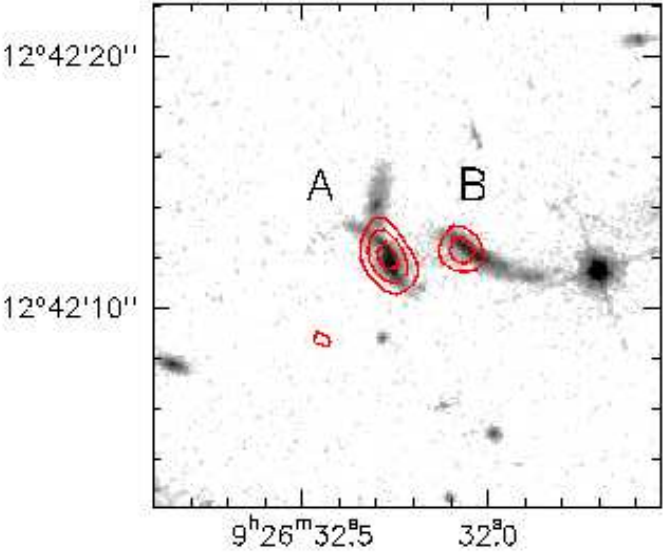}}
\caption{Contours of the two detections in CL 0926+1242 superimposed on the HST ACS/F814W image. We applied a small shift of 0.4'' in RA and 0.3'' in DEC between the CO and optical images, reflecting the GSC1 catalog uncertainties in absolute astrometry used by HST at the time
of the observations. The contours are at 50\%, 70\%, and 90\% of the peak maximum (1.6 Jy.km/s). }
\label{hst-co-cl0926+1242}
\end{figure}

\begin{figure}
\centerline{
\includegraphics[angle=-90, scale=0.4]{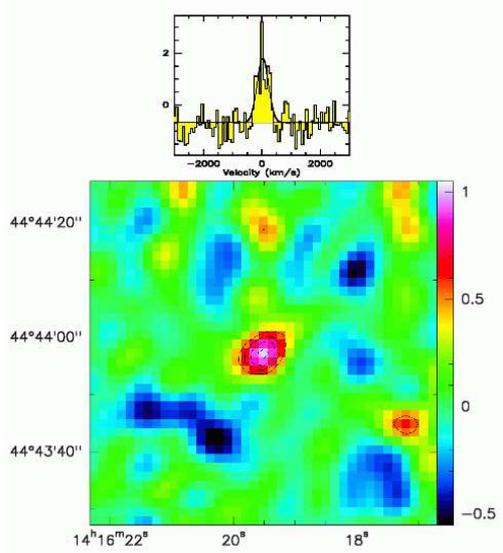}}
\caption{Intensity map and spectrum of GAL1416+4446. The color wedge of the intensity map is in Jy km/s, while
the spectrum shows the S$_{\mathrm{CO}}$ flux in mJy.}
\label{detection1416+4446}
\end{figure}


Table \ref{table:results} lists the CO fluxes of our three detected galaxies. We
assumed a ratio of 1 between the CO(2-1) and CO(1-0) luminosities as expected for a
thermalized optically thick CO emission and therefore the line luminosities were
calculated as L$^{\prime}$$_{\mathrm{CO}}$ = 3.25 10$^7$ S$_{\mathrm{CO}}\Delta v$
$\nu_{\mathrm{obs}}^{-2}$ (1+z)$^{-3}$ D$_L^2$ in K km s$^{-1}$ pc$^2$, where CO is
CO(1-0) for CL 1416+4446 and CO(2-1) for CL 0926+1242, D$_L$ the luminosity distance
in Mpc, z the redshift of the source, S$_{\mathrm{CO}}$ $\Delta v$ the total velocity
integrated line flux in Jy km/s, and $\nu_{\mathrm{obs}}$ the observed frequency
\citep{SolomonVandenBout} .

Figure \ref{LIR-LCO} shows L$^{\prime}$$_{\mathrm{CO}}$ and L$_{\mathrm{IR}}$ for our target galaxies
compared to local cluster galaxies.  Figure \ref{LIR-LCO} also shows
five galaxies of \citep{geach2011} in the outskirts of the rich cluster Cl 0024+16
($z$ = 0.395), normal BzK $z$$>$1 field galaxies \citep{daddi2010}, the serendipitous
detection of a $z$=1.114 galaxy in the cluster ISCS J1432.4+3332 \citep{wagg2012},
and the eight $z$$\sim$1.2 and eight $z$$\sim$2.3 galaxies with CO detections from
\citet{tacconi2010}. We derived the IR luminosity of the sample of
\citet{tacconi2010} by assuming a Kennicutt relation between L$_{\mathrm{IR}}$ and
star formation rate, rescaled for a Chabrier IMF, as in \citet{nordon2010},
SFR$_{\mathrm{IR}}$(M$_\odot$/yr) = L$_{\mathrm{IR}}$/9.85 × 10$^9$ L$_\odot$. We restrict our discussion to
samples of LIRGs, i.e., do not consider ULIRGs.

\begin{table}[t!]
\caption{Results from PdBI}             
\label{table:results}      
\centering                          
\begin{tabular}{l c c  c c} 
\hline
ID         & CO & $\Delta$V   & S$_{\mathrm{CO}}$        & L$^{\prime}_{\mathrm{CO}}$ \\ 
               &      & km s$^{-1}$    & Jy km/s     & 10$^{9}$K km/s pc$^2$   \\ 
\hline\hline               
GAL1416+4446   &  (1-0) & 420$\pm$40   &   1.0$\pm$0.1  &  8.1$\pm$0.8 \\
GAL0926+1242-A &  (2-1) & 420$\pm$40   &   0.6$\pm$0.1 &  1.9$\pm$0.3 \\
GAL0926+1242-B & (2-1) &  200$\pm$20   &   0.5$\pm$0.1 &  1.6 $\pm$0.3 \\
\hline
\end{tabular}
\end{table}
One immediate result from Figure \ref{LIR-LCO} is that the relation between L$_{\mathrm{IR}}$ and
L$^{\prime}$$_{\mathrm{CO}}$ for distant cluster galaxies extends the relation of
nearby systems to larger $L_{IR}$. The evolution of the bright end of the
IR cluster luminosity function was also noticed by \citet{bai2009} and
\citet{finn2010} and is not a consequence of our selection of cluster galaxies
with both CO and IR measurements. Nevertheless, there is no hint for any deviation
from the local trend such as seen for comparable redshift range by
\citet{combes2011, combes2013} when they include field ULIRGs; all points fall within
the 0.28dex dispersion of the local relation. Another interesting feature is that
the smallest CO luminosities among the distant galaxies are measured in cluster
galaxies.

The masses of our sample galaxies are shown in Figure \ref{LCOMstar}. The mass of
GAL0926+1242 B is very uncertain due to its faintness and correlated large
photometric errors. Still it is likely the least massive LIRGs observed so far in CO,
all the other distant galaxies being on the massive tail of the galaxy mass
distribution. In Section \ref{local} we showed that stellar masses play a secondary
role in determining star formation activity; comparing the nearby and distant samples
shows the lack of evolution with time of the conversion between gas and stars.
Nevertheless, we also saw that at fixed L$_{\mathrm{IR}}$, the field galaxies could
possess more cold gas than the cluster ones. This suggests that
L$^{\prime}$$_{\mathrm{CO}}$ mostly derived from the CO(1$-$0) line is sensitive to
the vast majority of the molecular gas in the cold interstellar medium of galaxies,
but less to the part which will actually be used to form stars. Higher order
transitions, from CO(4$-$3) to CO(7$-$6), or the HCN and CS molecules, have been
shown to trace gas that is denser and more directly connected to star formation
\citep{gao2004, bayet2009, bayet2009b}. This suggests that some molecular gas can be
stripped for example by ram-pressure or strangulation before the star formation
rate is affected. After a while, though, the fueling of the cluster galaxy activity
will be low and this might explain the systematic lower fraction of IR emitters in
the clusters as compared to the field.

\begin{figure} [t!]
\includegraphics[scale=0.45]{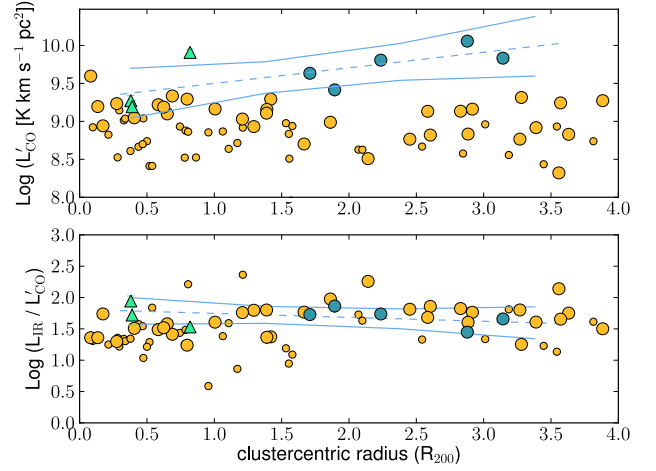}
\caption{Upper panel: The relation between the galaxy L$^{\prime}$$_{\mathrm{CO}}$
  and its projected location within the parent cluster expressed as fraction of the
  cluster virial radius R$_{200}$. The color code is the same as in Figure
    \ref{LIR-LCO}.  We highlight with larger orange symbols the local galaxies with
    2.5$\times$10$^{10}$L$_{\odot}$ $\leq$ L$_{\mathrm{IR}}$$\leq$
    10$^{11}$L$_{\odot}$ (see text).  Lower panel: The variation of the ratio
  between the IR and CO luminosities, respectively taken as proxy for star formation
  activity and cold gas reservoir, as a function of the distance to the cluster
  center.  The error bars on the PdBI measurements are plotted on the y-axes, however
  they are often smaller than the symbols. In both panels, the dash line is the
  linear regression fit of the relations for the intermediate redhift cluster
    galaxies and the plain lines are the 95\% confidence limits.}
\label{r-LCO}
\end{figure}

Figure \ref{r-LCO} compares the clustercentric distances and
L$^{\prime}$$_{\mathrm{CO}}$ for our galaxies and those of \citet{geach2011}.  We
used the location of the BCG (J2000, 00:26:35.7, +17:09:43) as the center of
CL0024+16 \citep{treu2003} and X-ray centers for CL0926+1242 and CL1416+4446
\citep{burenin2007}.  Figure \ref{r-LCO} shows a decline of
L$^{\prime}$$_{\mathrm{CO}}$ towards the center of the clusters, suggesting that
H$_2$ is depleted in the inner regions. The slope of the relation is 0.20 $\pm$ 0.09
with a correlation coefficient of 0.70 and a p- value of 0.05, rejecting the null
hypothesis. The lower panel of Fig. \ref{r-LCO} shows how the ratio between the IR
flux (proxy for the star formation) is linked to the CO luminosity, representative of
the amount of cold gas. We find no correlation (slope =$-0.06 \pm 0.05$). These
  galaxies have similar IR luminosities ( a factor 4 excursion, in principle leading to doubling
  L$^{\prime}$$_{\mathrm{CO}}$, from Fig.\ref{LIR-LCO}) and star formation
  efficiencies, therefore, they give evidence of a genuine enhancement in CO content
  towards the cluster outskirts.

For comparison, we show the same relations for the local sample. Because both
  our observations and those of Geach et al. targeted relatively massive clusters,
  we restrict this comparison to nearby galaxy structures with $\sigma
  \geq$300km/s. Furthermore, in order to minimize the aperture corrections and to
  allow for a large radial coverage, we consider z $\geq$0.01.  Following
  \citet{finn2010}, R$_{200}$ were calculated from the cluster velocity
  dispersions taken from \citet{struble1999, koranyi2002} and \citet{bothun1983}.
  The 0.30 dex dispersion in L$^{\prime}$$_{\mathrm{CO}}$ reflects the large range
  of IR luminosity covered, a standard variation of $\sim$1.7dex for a median
  value L$_{\mathrm{IR}}$=3.3$\times$10$^{10}$L$_{\odot}$.  Therefore, in order to
  allow for a fair comparison with the z$\sim$0.4 sample, we highlight the local
  galaxies with L$_{\mathrm{IR}}$ from 2.5$\times$10$^{10}$L$_{\odot}$ to
  10$^{11}$L$_{\odot}$, i.e., with the closest possible star formation rates to
  the Geach's and our PdBI galaxies, and spanning a similar factor 4 in IR
  luminosity. Contrary to the case of the z$\sim$0.4 clusters, we find no
  variation of the galaxy CO luminosity with the position of the galaxy. This
  absence of trend remains if one chooses a lower central L$_{\mathrm{IR}}$. The
  interpretation of the different behavior between the mid-z and nearby cluster
  galaxies is not straightforward, particularly because they belong to different
  star formation rate (L$_{\mathrm{IR}}$) regimes. The fact that most of the local
  sample are built from structures located in massive superclusters
  (Perseus-Pisces, Coma) could be one, added to the longer time evolution. In any
  case, it emphasizes the significance of the radial trend detected in the mid-z
  clusters.

We could question our choice of not using any conversion factor from
L$^{\prime}$$_{\mathrm{CO}}$ to a proper H$_2$ mass and suspect that we are only
witnessing different factors rather than a genuine change in cold gas reservoir.
The conversion ratio is known to vary significantly from galaxy to galaxy, and is
the lowest in active starburst galaxies, the ULIRGs, where a factor of $\alpha$=
0.8 (K km s$^{-1}$ pc$^2$)$^{-1}$ has been advocated by \citet{solomon1997}. The
conversion factor is depending on the excitation temperature of the CO lines,
T$_{\mathrm{ex}}$, the density of the molecular gas, n$_{\mathrm{H2}}$, the gas
metallicity Z, and the ambient UV radiation field, among others. It is expected to
vary as n$_{\mathrm{H2}}$$^{1/2}$/T$_{\mathrm{ex}}$, from the Virial hypothesis
for molecular clouds \citep[e.g.,][]{dickman1986}.  When the gas is more excited
and hotter, due to an intense and concentrated starburst as in ULIRGs, the
conversion ratio decreases.  The dependence in metallicity is complex and
non-linear: at low metallicity, C and O being less abundant, yields a lower
CO/H$_2$ abundance, but also, there is then less dust, and less self-shielding
from the UV photodissociating radiation, so that CO is preferentially destroyed,
before H$_2$ itself \citep{maloney1988}.  Given the high mass of our galaxies, the
metallicity should be at least solar or above, and this parameter should not be
determinant on the CO-to-H$_2$ conversion ratio. Our galaxies are only LIRGs and
at their intermediate redshifts, they correspond to main sequence galaxies,
instead of the starburst extreme of the star formation activity \citep{finn2010,
  wuyts2011}.  Therefore a standard and constant conversion ratio most likely
applies to our galaxies. This should of course be confirmed through more
observations, for instance of the dust mass, and dust-to-gas mass ratio
\citep{sandstrom2012}.

\section{Conclusions}
\label{conclusion}

We used IRAM PdBI to obtain the first CO detections of galaxies located at projected
distances within the virial radii of two intermediate redshift clusters, CL1416+4446
(z=0.397) and CL0926+1242 (z=0.489). We detected three galaxies at high significance
(5-10$\sigma$) enabling robust estimates of their CO luminosities.  All three
galaxies are LIRGs, and they are normal star forming galaxies.  The two galaxies in
CL0926+1242 are interacting spirals. In order to put our results into a general
context, we revisited the case of the nearby galaxies ($z\geq$0.05).  We assembled a
sample of 275 field and 170 cluster galaxies for which we could collect both reliable
far infrared luminosities and CO luminosities from the literature. We revised the
distances of all galaxies, derived the far infrared luminosities and
aperture-corrected the CO luminosities in a homogeneous way. Our main conclusions are
the following:

$\bullet$ At fixed L$_{\mathrm{IR}}$, or fixed stellar mass, the frequency of a high
L$^{\prime}${$_{\mathrm{CO}}$ is lower in clusters than in the field, suggesting
  environmental depletion of the reservoir of cold gas.

$\bullet$ The level of star formation activity in a galaxy is primarily linked to the
  amount of cold gas, rather than to the galaxy mass or redshift. In clusters, just
  as in the field, the conversion between gas and star looks universal. The relation
  between L$_{\mathrm{IR}}$ and L$^{\prime}$$_{\mathrm{CO}}$ for distant cluster
  galaxies extends the relation of nearby systems to higher IR luminosities; the
  distant galaxies fall well within the dispersion of the trend defined by local
  systems.

$\bullet$ Given that L$^{\prime}$$_{\mathrm{CO}}$ is mostly derived from the
  CO(1$-$0) line and sensitive to the vast majority of the molecular gas in the cold
  interstellar medium of galaxies, but less to the part which will actually be used
  to form stars, we suggest that molecular gas can be stripped before the star
  formation rate is affected. After a while, though, the fueling of the cluster
  galaxy activity will be low and this might explain the systematic lower fraction of
  IR emitters in the clusters as compared to the field. If this is true, for a given
  star formation rate, field and cluster galaxies should have comparable mass of
  dense gas as traced by the high order transitions of CO, or the HCN and CS
  molecules.

$\bullet$ Combining the sample of \citet{geach2009,geach2011} and ours, we find
  evidence for a decrease in CO towards the cluster centers. This is the first
  hint of an environmental impact on cold gas at intermediate redshift.

Our present work is by no means a definitive answer to the questions
raised in the introduction. Instead it stresses the imperative need for large
samples of LIRGs both in the field and in clusters,  to put studies such as this one
on firm statistical ground.  Compared to local surveys, surveys at intermediate
redshift do not suffer large luminosity uncertainties due to partial spatial coverage.
Nevertheless, distant surveys must account for: {\it i)} chance alignment:
galaxies in the outskirts of the clusters have a higher probability to be
genuinely located in the infall regions than have galaxies projected on the
cluster core to be indeed in the inner regions. {\it ii)} diversity in cluster
properties: the mass of the clusters largely determines both the proportion of
galaxies with ongoing star formation, and the strength of their activity, as
infered from the [OII] emission line \citep[e.g.,][]{poggianti2006}

\bibliographystyle{bibtex/aa}
\bibliography{coldgas.r2}

\acknowledgement{This work was supported by the Swiss National Science Foundation
  (SNSF). FC acknowledges the European Research Council for the Advanced Grant
  Program Num 267399-Momentum. KR was funded by a Cottrell College Science Award from
  the Research Corporation.  We are very grateful to Dr C.S. Chang at IRAM for the
  help provided during the data calibration and reduction.  This research has made
  use of the NASA/ IPAC Infrared Science Archive, and the Extragalactic Database
  (NED), which are operated by the Jet Propulsion Laboratory, California Institute of
  Technology, under contract with the National Aeronautics and Space Administration.
  It makes use of data products from the Wide-field Infrared Survey
  Explorer, which is a joint project of the University of California, Los Angeles,
  and the Jet Propulsion Laboratory/California Institute of Technology, funded by the
  National Aeronautics and Space Administration and is based in part on
  observations made with the Spitzer Space Telescope, which is operated by the Jet
  Propulsion Laboratory, California Institute of Technology under a contract with
  NASA.}

\end{document}